\documentstyle[aps]{revtex}
\begin{document}
\title{Novel criticality in a model with absorbing states\\}
\author{Adam  Lipowski \cite{byline}\\}
\address{Department of Physics, A. Mickiewicz University, Ul. Umultowska 85, 
61-614 Pozna\'{n}, Poland}
\date{\today}
\maketitle
\begin{abstract}
We study a one-dimensional model which undergoes a transition
between an active and an absorbing phase.
Monte Carlo simulations supported by some additional arguments prompted as
to predict the exact location of the critical point and critical exponents in
this model.
The exponents $\delta=0.5$ and $z=2$ follows from random-walk-type arguments.
The exponents $\beta = \nu_{\perp}$ are found to be non-universal and encoded in the singular
part of reactivation probability, as recently discussed by H.~Hinrichsen (cond-mat/0008179).
A related model with quenched randomness is also studied.
\end{abstract}
\section{Introduction}
Recently, non-equilibrium phase transitions have been intensively studied in 
variety of models.
Some attempts were also made to classify possible types of phase
transitions into universality classes similarly to equilibrium
transitions~\cite{HAYE1,EVANS}.
However, non-equilibrium transitions are much reacher and puzzling than
equilibrium ones and their understanding is still far from complete.

A class of nonequlibrium models for which such a categorization seems
most feasible are models with absorbing states~\cite{DICK}.
For example, models with unique absorbing states
 most likely belong to the
so-called directed-percolation (DP) universality
 class~\cite{GRASSJAN}.
Moreover, models with double (symmetric) absorbing
states or with some conservation law in their dynamics belong to
another universality class, the so-called parity-conserving (PC) universality
class ~\cite{PARITY}.
Although the above classification is best confirmed for one-dimensional
models, it seems to apply to higher-dimensional models as well~\cite{COMM1}.
Moreover, models with more than two~\cite{JANSSEN} or even infinitely many
absorbing states also fall into the above universality classes. (Typically, models
with infinitely many absorbing states fall into DP universality class,
however, certain symmetries~\cite{PARK,LIP2000} or conservation
laws~\cite{MENDES} might change the criticality into PC.)

In the present paper we study certain one-dimensional models with
 infinitely
many absorbing states.
In the first part we study a model defined on a ladder-like lattice (ladder model) and our Monte Carlo
simulations show that this model belongs to neither DP nor PC universality classes.
Critical exponents $\delta$ and $z$ of this model appear to be 
rational numbers (contrary to the DP or PC criticality) and can be obtained using simple
random-walk-type arguments.
However, the exponents $\beta$ and $\nu_{\perp}(=\beta)$ appears to be non-universal.
It turns out that their values are encoded in the singular part of the reactivation probability as
was recently postulated in the context of related models by Hinrichsen~\cite{HINNEW}.

Using certain property of our ladder model we can easily calculate the upper bound on the existence
of the absorbing phase.
Numerical simulations show that this bound most likely give exact location of the critical
point in this model.
In the final part of our paper we examine yet another but closely related model (single-chain
model) for which we can obtain also the lower bound of the absorbing phase.
It turns out that both bounds are the same which exactly locates the transition point in
this model.
The model studied in this part of our paper contains some quenched random variables and
its critical
behaviour is similar to that of the model studied in the main part of our paper.
\section{Ladder Model}
\subsection{Definition and basic properties}
Our model is a certain variant of models recently studied in some other
contexts~\cite{LIP2000,LIPLOP}.
It is defined on a one-dimensional ladder-like lattice (see Fig.~\ref{f1}a).
For each bond between the nearest-neighbouring sites we introduce
a bond variable 
$w\in (-0.5,0.5)$.
Introducing real parameters $r$ and $s>0$, we call a given site active when 
$w_1w_2|w_3|^s<r$, where $w_1$ and $w_2$ are intra-chain bond variables connected
with this site and $w_3$ is the inter-chain bond variable (see Fig.~\ref{f1}a).
Otherwise, this site is called non-active.
The model is driven by random sequential dynamics and when the
active site is selected, we assign anew, with uniform probability, three
neighbouring bond variables.
Nonactive sites are not updated.

First, let us notice that the model has a certain global up-down symmetry.
Namely, reversing all bond variables $w\rightarrow -w$ we do not change the
status of any site, i.e., active (non-active) sites remains active
(non-active). The same symmetry appears in a closely related model which
is defined on a single chain and where activity of a site is determined by the
product of two intra-chain variables~\cite{LIP2000} (in the following we refer
to this model as model A).
It was suggested that this symmetry is responsible for the fact that the
model A belongs to the PC universality class~\cite{LIP2000}.
If so, the present model should also belong to this universality class.
In the present paper we show, however, that the critical behaviour of this
model is different.
In our opinion, this result does not completely abolish the claims about the
importance of the up-down symmetry.
Namely, as we will show below, the change of the universality class in the present
model comes from the very intricate mechanism: upon approaching the critical point the
system might be regarded as composed of two weakly-interacting critical
subsystems.
We expect that in a generic case such a mechanism is absent and  models of this
kind which are endowed with the up-down symmetry should belong to the PC
universality class.

This situation bears some similarity to the equilibrium statistical mechanics,
where one expects that short-range interacting Ising models will generically
belong to the Onsager universality class. 
It is known, however, that when certain additional interactions are included,
the critical behaviour of an Ising model, which is then equivalent to the
eight-vertex model, deviates from the generic case~\cite{BAXTER}.
\subsection{Monte Carlo simulations}
To examine the properties of the above model we used standard Monte Carlo
simulations.
First, let us describe results for $s=2$.
We calculated the steady-state density of active sites $\rho$.
The fact that the model is defined on a one-dimensional lattice enabled us
to examine systems of large size $L$ (up to $5\cdot 10^5$).
Also the simulation time $t_{{\rm MC}}$ was rather large and typically
$t_{{\rm MC}} \sim 10^5-10^6$ (the unit of time is defined as a single, on
average, update per lattice site). 
Our simulations show that for $r>0$ the
density $\rho$ remains positive but as soon as $r$ becomes negative the system
quickly reaches one of the absorbing states and thus $\rho=0$. A simple
argument, similar to the one used for related models~\cite{LIP2000}, shows
that for $r<0$ the model gradually generate sites which remains permanently
non-active. Indeed, let $r<0$ and a certain inter-chain bond $w_3$ satisfies
the condition
\begin{equation}
|w_3|<\sqrt -r/(0.5),
\label{e1}
\end{equation}
then for each site connected to $w_3$ the absolute
value of the product $w_1w_2w_3^2$ is smaller than $-r$, which implies
$w_1w_2w_3^2>r$ and these sites remain permanently non-active.
Since upon updating an active site there is a finite probability to satisfy
(\ref{e1}), there is a finite rate of creation of such sites (a similar argument can be
applied to an intra-chain bond).
Thus, in agreement with numerical simulations, for $r<0$ the system is in the
absorbing phase.
Combined with the computationally observed property that for $r>0$ the density
$\rho$ remains positive, it implies that $r=r_{{\rm c}}=0$ is the transition
point for that model which separates active and absorbing phases.
In addition, our simulations, which are described below, show that $r=0$ is actually a
critical point of this model and is accompanied by typical power-law characteristics.

The first evidence of such a characteristic is shown in Fig.~\ref{ro}, which
presents the logarithmic plot of $\rho$ as a function of $r$.
The linearity of the plot for approximately three decades confirms the
power-law behaviour $\rho\sim r^{\beta}$.
At the same time, it confirms the location of the critical point $r_{{\rm c}}=0$.
From the least-square analysis of these data we estimate $\beta=0.51(1)$.

Another quantity which we measured was the time evolution of the density
$\rho$ for $r$ close to the critical point $r=r_{{\rm c}}$.
One expects that at criticality $\rho\sim t^{-\delta}$ and deviations from the
power-law behaviour appear off the critical point.
The results are presented in Fig.~\ref{time}. 
For $r=0$ the clear power-law behaviour is observed and from the slope of the
data we estimate $\delta=0.50(1)$.

Other power-law characteristics are obtained from the finite-size analysis.
In Fig.~\ref{sizetau} we present the size dependence of the characteristic time
$\tau$ defined as an average time needed for the system to reach an absorbing
state (with a random initial configuration).
Again, the best linearity is seen for $r=0$ and in this case we obtain $\tau\sim
L^z$ with $z=2.01(1)$.
Let us notice that this figure limits the allowed values of $r_{{\rm c}}$ to the
very narrow range $-10^{-12}<r_{{\rm c}}<10^{-8}$.

Finally, Fig~\ref{sizero} presents the steady-state density $\rho$ as a function
of size $L$.
For $r=10^{-4}$ and $10^{-6}$, $\rho$ converges to the positive value but for
$r=0$ it decays as $\rho\sim L^{-\beta/\nu_{\perp}}$ with the exponent
$\beta/\nu_{\perp}=0.99(1)$.

For comparison with our model, we quote the values of these exponents for the
one-dimensional DP~\cite{JENSEN} and PC~\cite{JENSEN1} universality classes:
$\beta=0.26486$(DP), 0.92(PC); $\delta=0.159464$(DP), 0.286(PC);
$z=1.580745$(DP), 1.74(PC) and $\nu_{\perp}=1.096854$(DP), 1.83(PC). 
Thus, our results clearly places the model into a new universality class.

However, the obtained values of the exponents are not entirely unexpected.
In the following we argue that at least the values of $\delta,\ z$  and of the
ratio $\beta/\nu_{\perp}$ can be inferred from the properties of some other
models. 
First, let us examine in more detail the model at the criticality, i.e., at
$r=0$. Let us notice, that in this case it is only the sign of the expression
$w_1w_2w_3^2$ which determines the state of a site.
Since $w_3^2$ is always positive, it means that at the criticality this term is
irrelevant and the state of a site is determined by the product $w_1w_2$, i.e.,
by the product of intra-chain variables.
The same condition appears in model A and
thus the present model at the criticality is equivalent to two non-interacting
models A.  Numerical evidence was already presented~\cite{LIP2000} that at
$r=0$ model A is at the end-point of its critical phase which in A model
appears in the range $0<r<r_{{\rm c}}\sim 0.027$. This critical phase is
described by simple, random-walk related exponents: $\delta=0.5,\ z=2$ and
$\beta/\nu_{\perp}=1$~\cite{ALEMANY}. 
On the basis of the above relation and in agreement with our simulations we
obtain that these random-walk exponents are the critical exponents in our
model. Let us emphasize that the critical phase in PC models exists in a
certain range of a control parameter~\cite{COMM2} and its (random-walk)
criticality is described by a different set of exponents than the critical
point of such models. In our model, there is only an isolated critical point
which is described by the random-walk exponents.

The above relation with model A does not explain the value of the
exponent $\beta$, which most likely equals 0.5.
This is because the relation with A model holds only at $r=0$ while the
exponent $\beta$ describes the off-critical singularity.
One can argue, however, that when $r$ is positive but small, then the
interaction between these two models A is also small.
Let us recall now, that for such $r$ model A is critical and thus the present
model is equivalent to two weakly-interacting critical models A.
It is interesting to observe that such an interaction is sufficient to destroy
the criticality and to keep the system in the active phase with $\rho>0$.
Qualitatively, we explain the large (comparing to model A) activity as follows:
In models with absorbing states a non-active domain can be reactivated only through its
boundaries.
However, the already mentioned weak interaction can reactivate even the interior of
such domains, which dramatically increases activity of the system.
\subsection{Reactivation probability}
In a recent paper Hinrichsen has shown that for certain models the singularity of the order
parameter is determined by singularity of the reactivation probability $W$~\cite{HINNEW}.
His argument relies on the observation that $W$ usually scales linearly with
the control parameter (i.e., a parameter which measures the distance from a critical point).
Thus, in models where the density of active sites $\rho$ scales linearly with $W$, the singularity of
$\rho$ as a function of the control parameter is solely a consequence
of the singular behaviour of $W$.
In this subsection we show that Hinrichsen's results extends also to the model examined in
this paper.

Let us define the reactivation probability $W(r)$ as a probability that a given active site
remains active after an update.
To compute $W(r)$ we first calculate the probability density $P_s(z)$ that $w_1w_2|w_3|^s=z$
where $w_1,\ w_2$ and $w_3$ are independent and uniformly distributed on $(-0.5,0.5)$.
Thus, we have to calculate
\begin{equation}
P_s(z)=\int_{-1/2}^{1/2}dw_1\int_{-1/2}^{1/2}dw_2\int_{0}^{(1/2)^s}\frac{2}{s}\tilde{w}_3^{1/s-1} 
\delta(z-w_1w_2\tilde{w}_3)d\tilde{w}_3,
\label{2}
\end{equation}
where $\frac{2}{s}\tilde{w}_3^{1/s-1}$ is the probability density of $\tilde{w}_3=|w_3|^s$, where $w_3$
is uniformly distributed on $(-1/2,1/2)$.
The calculation of the integrals (\ref{2}) is elementary.
Perfoming integration over $w_1$, $w_2$ and then over $\tilde{w}_3$ we obtain
\begin{equation}
P_s(z)=
\frac{-4}{s}\int_{0}^{(1/2)^s}\tilde{w}_3^{1/s-2}\ln(4|z|/\tilde{w}_3)\Theta
(\tilde{w}_3-4|z|)d\tilde{w}_3= 
\frac{4}{s-1}\{2^{s-1}[\ln (4|z|)+s\ln(2)+\frac{s}{1-s}]+\frac{s}{s-1}(4|z|)^{1/s-1}\},
\label{3}
\end{equation} 
where $\Theta(x)$ is the unit step function.
Having calculated $P_s(z)$, the reactivation probability $W(r)$ for positive $r$ is given as
\begin{equation}
W(r)=\int_{(1/2)^{s+2}}^{r}P_s(z)dz=\frac{1}{2}+\frac{4}{s-1}\{ 2^{s-1}
[r(\ln(4r)-1)+sr(\ln(2)+\frac{1}{1-s})]+\frac{s^2}{4(s-1)}(4r)^{1/s}\}.
\label{4}
\end{equation}
From (\ref{4}) one can see that $W(r)$ is indeed singular at the critical point $r=0$ and for 
$s>1$ $W(r)\sim r^{1/s}$.
One can see that the numerically found singularity of the order
parameter with the exponent $\beta=1/2$ for $s=2$ corresponds to the same singularity of $W(r)$.
Thus, the density $\rho$ in the vicinity of the critical point $r=0$ should scale linearly with the
reactivation probability $W(r)$ and the singular behaviour is only due to the singular behaviour of
$W(r)$ as a function of $r$.

To check the above arguments we performed simulations for $s=4$ and 1/2.
Let us notice that for $s=1/2$ the leading term is not $r^{1/s}$ but $r\ln (4r)$.
In this case $\beta$ should be unity but the logarithmic corrections might substantially affect the
scaling behaviour. 
The results are presented in Fig.~\ref{ro}.
As estimated from the slope of our data, $\beta=0.26(1)$ for $s=4$ and $\beta=1.3(2)$ for $s=1/2$.
For $s=4$ the exponent $\beta$ is very close to 1/4 which clearly confirms our arguments.
For $s=0.5$ our data are less accurate mainly due to very large relaxation time.
Moreover, the logarithmic corrections might be responsible for the fact that the 'true' scaling is not
clearly seen.

Let us also notice that at the critical point $r=0$ only the sign matters and the behaviour of the
ladder model is independent on $s$.
Thus, the remaining exponents $\delta,\ z$ and the ratio $\beta/\nu_{\perp}$ must be the same as in the
$s=2$ case.
\section{Single-Chain Model with Quenched Disorder}
In this section we examine a simple model which might help us to understand the behaviour of
the ladder model studied in the previous section.
The model is defined on a single chain (see Fig.~\ref{f1}b).
In addition to bond variables $-1/2<w<1/2$ there are on-site quenched variables $0<v<1$.
A given site is defined as active if $w_1w_2v<r$, where $w_1$ and $w_2$ are variables on bonds
connected to this site and $v$ is the corresponding site variable.
Once initially selected, $v$-variables remain unchanged during the evolution of this model
(evolving variables are only bond variables). 
(The already discussed model A corresponds to the case $v=1$ on each site.)

It is elementary to prove the following properties of the this model:\\
(i) Similarly to the model studied in the previous section the model for $r<0$ generates with a
finite rate sites which remain permanently non-active.
It means that for $r<0$ the model is in the absorbing phase.\\
(ii) Let $r>0$.
Let us consider a site for which $v<4r$.
It is obvious that for any choice of bond variables attached to it this site remains permanently
active.
Since  in the thermodynamic limit a finite fraction $4r$ of sites has $v<4r$, we obtain that for
$r>0$ the model is in the active phase.

As a conclusion, we obtain that $r=0$ separates active and absorbing phases of this model.
Let us also notice that for $r=0$ the model is equivalent to the model A (i.e., $v$ variables are
irrelevant), which was shown to be critical at that point~\cite{LIP2000}.
Moreover, the fraction $4r$ of permanently active sites decreases linearly to zero which suggests
that in this model $\beta=1$~\cite{COMM3}.

Monte Carlo simulations of this model strongly suggest that indeed $\beta=1$ (see~Fig.~\ref{ro}).
Similarly to the ladder model, at $r=0$ only the sign matters and $v$-variables are irrelevant.
Thus, at criticality the behaviour of this model is the same as of the A model at $r=0$.
Numerical simulations of the latter model show~\cite{LIP2000} that in this case $z=2$ and $\delta=1/2$
as in the ladder model.
\section{Conclusions}
In conclusion, we examined a class of models with infinitely many absorbing states.
Critical exponents $\delta,\ z$ and $\beta/\nu_{\perp}$ take random-walk values.
However, the exponents $\beta$ and $\nu_{\perp}$ are non-universal.
This non-universality is related to the singular behaviour of the reactivation probability 
$W(r)$.
Thus, when as a control parameter of the model we choose $W(r)$ rather than $r$, the order
parameter would have the universal exponent $\beta=1$.
It would be interesting to examine this model using, for example, field-theory
methods~\cite{CARDY} and to check whether some other universality classes among
models with absorbing states are possible.
\acknowledgements I thank Haye Hinrichsen for very stimulating correspondence and the Department of
Mathematics of the Heriot-Watt University (Edinburgh, Scotland) for allocation of computer time.

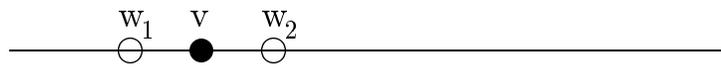
\begin{figure}
\caption{(a) The ladder model.
The site ($\bullet$) is active when $w_1w_2|w_3|^s<r$.
An update of an active site replaces all three neighbouring bonds ($\circ$)
with uniformly distributed random numbers from the interval (-0.5,0.5).
(b) The single-chain model.
The site is active when $w_1w_2v<r$ and its update replaces only $w_1$ and $w_2$.
} 
\label{f1} 
\end{figure} 
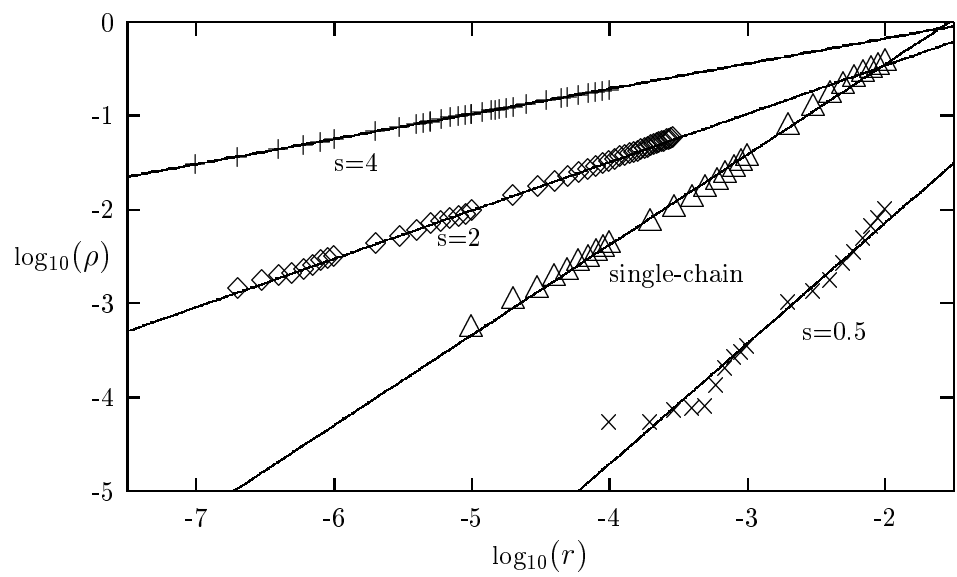
\begin{figure} 
\caption{The log-log plot of the density
of active sites $\rho$ as a function of $r$. The linear size $L$ in this
simulations varied from $L=5\cdot 10^4$ to $2\cdot 10^6$. 
The solid lines (least-square fit) have slopes
corresponding to $\beta=0.26(1)$ (for $s=4$), 0.51(1) (for $s=2$), 1.3(2) (for $s=0.5$) and
0.99(2) for the single-chain model.} 
\label{ro} 
\end{figure}  
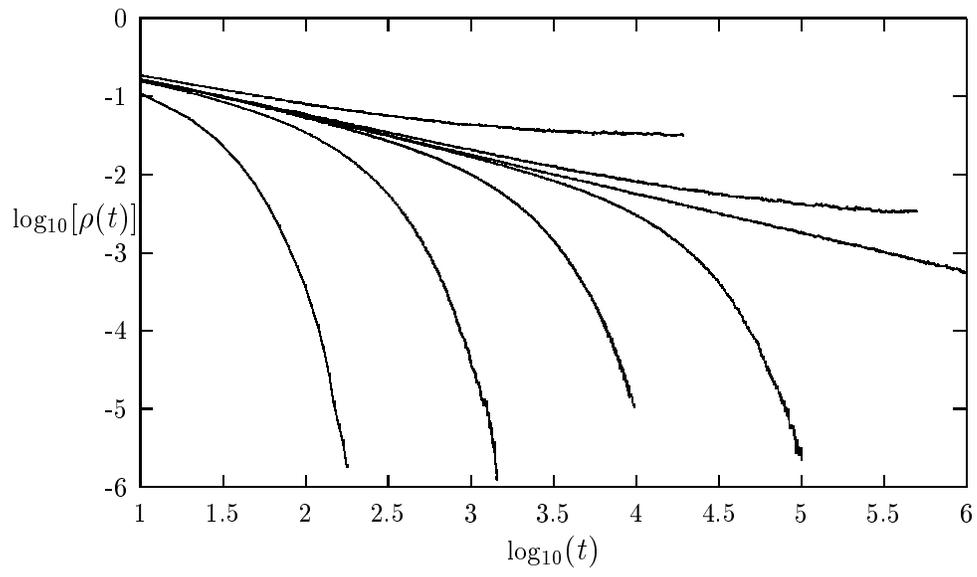
\begin{figure}
\caption{The time evolution of the density $\rho$ for (from top to bottom)
$r=10^{-4},\ 10^{-6},\ 0, \ -10^{-10},\ -10^{-8},\ -10^{-6},$ and $ -10^{-4}$.
The simulations were made for $s=2$ and $L=10^5$ and we checked that for the examined
time scale our data are basically size-independent. These data limit $r_{{\rm
c}}$ to the range $-10^{-10}<r_{{\rm c}}<10^{-6}$. Even tighter bounds follow
from Fig.~\protect\ref{sizetau}. } 
\label{time} 
\end{figure} 
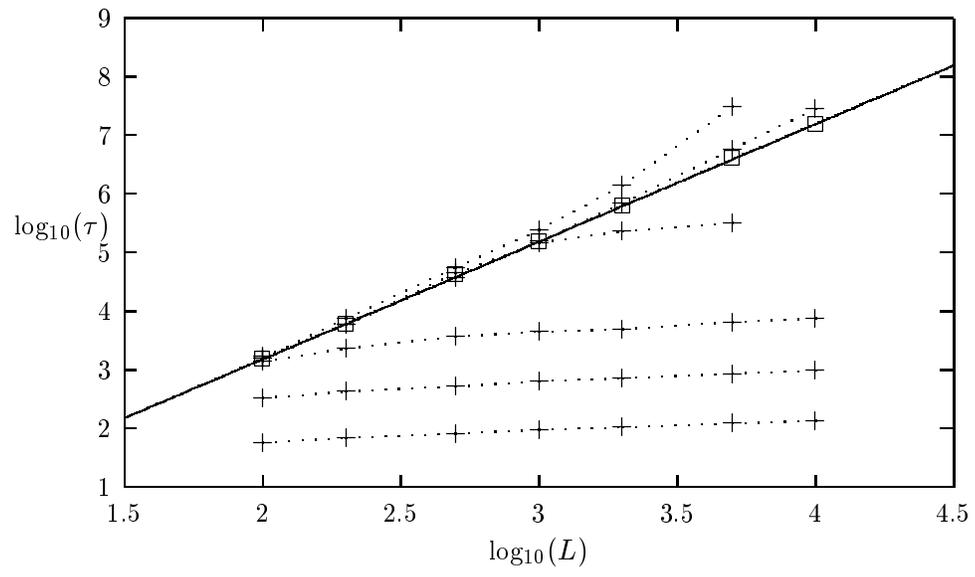
\begin{figure}
\caption{ The size dependence of the characteristic time $\tau$ for (from top
to bottom):   $r=10^{-6},10^{-8},0,-10^{-12},-10^{-8},\\ -10^{-6},$ and
$-10^{-4}$.
The straight line has a slope corresponding to $z=2.01$.
Each point is an average of 100 independent runs and $s=2$.} 
\label{sizetau} 
\end{figure} 
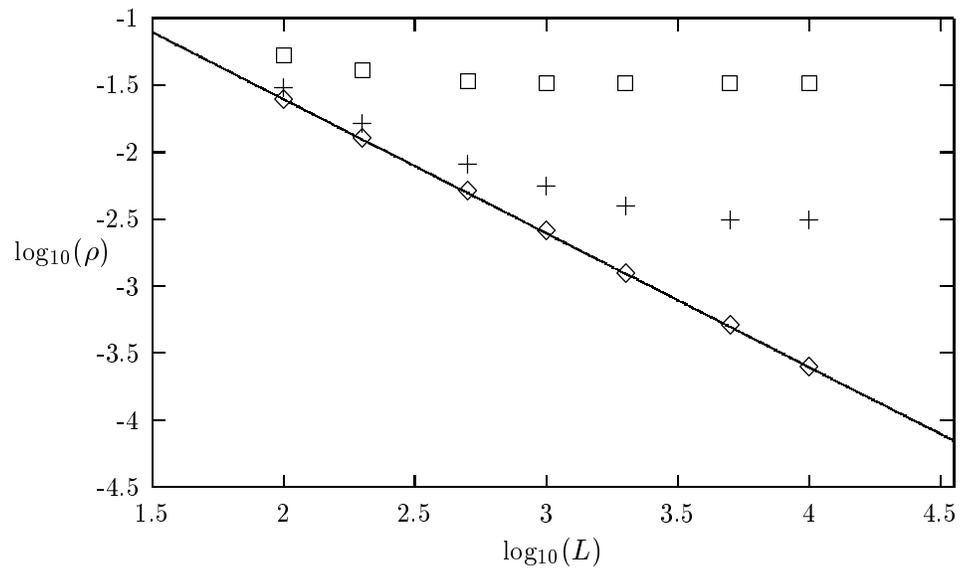
\begin{figure} 
\caption{The size
dependence of the steady-state density $\rho$ for $r=10^{-4} (\Box),10^{-6}
(+)$ and $0 (\Diamond)$. The straight line has a slope corresponding to
$\beta/\nu_{\perp}=0.99(2)$ ($s=2$). } 
\label{sizero}  
\end{figure}  

\begin{references}
\bibitem{COMM1} In higher dimensions PC models are divided into two distinct
universality classes (see, e.g.,~\cite{HAYE1} for details).
\bibitem{COMM2} Usually the entire non-active phase of PC models is critical.
The only example, of which we are aware of, where this phase terminates at a certain
value of a control parameter is model A.
\bibitem{COMM3} One might expect that for small but positive $r$ the activity is localized
only around these permanently active sites.
\end{references}
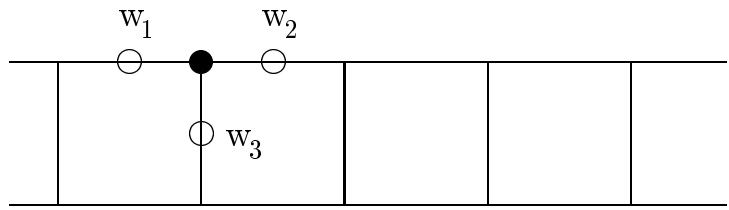
\end{document}